# Design and Comparative analysis of a Two-Stage Ultra-Low-Power Subthreshold Operational Amplifier in 180nm, 90nm, and 45nm technology

Sumukh Nitundil
Department of Electrical and Electronics Engineering
Birla Institute of Technology and Science, Pilani
Rajasthan, India
f20170375@pilani.bits-pilani.ac.in

Nihal Singh
Department of Electrical and Electronics Engineering
Birla Institute of Technology and Science, Pilani
Rajasthan, India
f20171159@pilani.bits-pilani.ac.in

Rushabha Balaji
Department of Electrical and Electronics Engineering
Birla Institute of Technology and Science, Pilani
Rajasthan, India
f20170220@pilani.bits-pilani.ac.in

Pankaj Arora
Department of Electrical and Electronics Engineering
Birla Institute of Technology and Science, Pilani
Rajasthan, India
pankaj.arora@pilani.bits-pilani.ac.in

*Abstract*— In this paper, a two-stage ultra-low-power operational amplifier is designed, and a comparative analysis of the proposed subthreshold complementary amplifier is presented between 180nm, 90nm, and 45nm CMOS technology. The proposed operational amplifier is compared across several different parameters to determine the optimal design. It achieves a maximum gain of around 75 dB and a phase margin of 76°, dissipating just 140nW with a supply voltage of 0.5 V which is well suited for biomedical applications that require low power and high gain. The proposed operational amplifier has been designed using a SPICE-based circuit simulator.

*Keywords—Ultra-low power, low voltage, operational amplifier, subthreshold operation, bio-amplifier*

## I. INTRODUCTION

Biomedical applications such as sensing of electrocardiogram (ECG) and electromyography (EMG) signals or implantable devices such as pacemakers and muscle stimulators, require low power and portable analog front-end systems [1]. These systems deal with the signals that are usually of low-frequency (less than a few kilohertz) with low amplitudes. Moreover, they are also slow signals which do not require a high slew rate [2].

In 2005, Wise group presented a 3-D microelectrode array to interface directly to cortical tissues for various applications involving neural processing [3]. This technology has been extensively used in the treatment of several neurological diseases [4] through medical implants and prostheses with an analog front-end for pre-processing the neural information. The work presented in this paper has been designed to optimize the required parameters such as low power and low rail-to-rail voltage. Spike identification and event recognition from detected signals require optimal noise reduction and limitation of output leads to manageable levels. This can be facilitated by an efficiently designed on-chip amplification and pre-processing circuitry as shown in Figure 1 consisting of several components such as operational amplifiers (bio-amplifier), low pass filters, and analog to digital converters of which the bio amplifier is the essential building block and main power consuming unit whose power needs to be minimized to increase the portability and battery life [5].

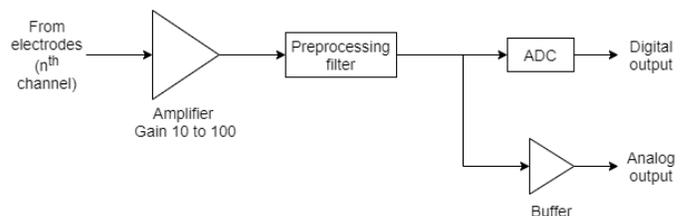

Fig. 1. The architecture of central nervous system using a microelectrode array

Subthreshold operation of MOSFETs uses lower voltages and lesser bias current, and hence categorized as ultra-low-power (sub $1\mu W$) [6]. The exponential dependence of the current generally leads to a higher gain as well. This makes it the ideal candidate for biomedical analog front end Operational Amplifier (OpAmp) design, with ultra-low power, high gain (due to small signal amplitude), low bandwidth and slew rate (due to the slower nature of signals). Advancement in VLSI scaling technology has led to the design of circuits in sub-micron size [7]. This further aids biomedical applications such as a pacemaker or cochlear implants that require small size and low power. However, increased technology scaling has led to short channel effects that degrade the performance of the circuit. Hence it has become crucial to analyze the performance of the circuits as they are scaled to different technologies.

In this work, an ultra-low-power and high gain subthreshold OpAmp is designed which is suitable for biomedical applications such as neural signal sensing and processing. A comparative analysis of the performance in 180nm, 90nm, and 45nm technologies is presented to give a comprehensive overview of the circuit performance with scaling, to enable smaller, lower power, yet reliable circuits.

## II. PROPOSED DESIGN

### A. Subthreshold operation

In the subthreshold region, the transistor operates below the threshold voltage, hence allowing the circuit to operate at the lower voltages and bias currents. The equation for drain current in the subthreshold region is given by [8]

$$I_D = \frac{W}{L} I_0 \exp\frac{V_{GS}-V_{th}}{mV_T}(1-\exp\frac{-V_{DS}}{V_T}) \quad (1)$$
$$\approx \frac{W}{L} I_0 \exp\frac{V_{GS}-V_{th}}{mV_T} \quad (when: V_{DS} > 3V_T)$$

where $W/L$ is the transistor aspect ratio, $V_{GS}$ and $V_{DS}$ are the gate-to-source and drain-to-source voltages, $V_{th}$ is the threshold voltage, and $V_T = k_B T/q = 26mV$ is the thermal voltage at room temperature ($k_B$ is the Boltzmann constant, T is the absolute temperature and q is the elementary charge), $m$ is the subthreshold slope parameter which is technology dependent and usually between 1 and 2 and $I_0$ is a process parameter that also depends on temperature.

From (1) the output current is related exponentially with the input voltage rather than the quadratic relationship followed in the region of strong inversion. This increases the transconductance of the MOSFET thereby resulting in a higher gain, with characteristics similar to a BJT.

After considering body and drain induced barrier lowering (DIBL) effects, the threshold voltage $V_{th}$ is given by [9]

$$V_{th} = V_{th0} - \lambda_D V_{DS} - \lambda_B V_{BS} \text{ with } \lambda_D \text{ and } \lambda_B > 0 \quad (2)$$

where $V_{BS}$ is the body to source voltage, $V_{th0}$ is the threshold voltage at room temperature (for $V_{BS} = 0V$), and $\lambda_D$ and $\lambda_B$ are the DIBL and body effect coefficients respectively [10]. From (1) and (2), the equation for transconductance $g_m$ can be derived [11]

$$g_m = \frac{I_D}{mV_T} \quad (3)$$

$$\frac{W}{L} = \frac{I_D}{I_0} \exp\frac{V_{TH}-V_{GS}}{mV_T} \quad (4)$$

The used constants values are shown in Table I.

Table 1: TABLE OF CONSTANTS

|  | $V_{th0}$ (mV) | $I_0$ (nA) | m | $\lambda_D$ |
|---|---|---|---|---|
| NMOS | 366.2 | 288 | 1.20 | 2.23e-3 |
| PMOS | 390.6 | 74 | 1.35 | 2.23e-3 |

### B. Design Methodology of two-stage op-amp

Figure 2 represents the transistor-level configuration of the two-stage Miller-compensated OpAmp in which no particular technique exploiting the MOSFET body terminal is used ($V_{BS} = 0V$). In the proposed circuit, all MOSFETs operate in the subthreshold region, and the architecture is designed to amplify the input signal across two stages, the first being the differential stage and the second being the gain stage.

Table II shows the design goals for the two-stage OpAmp into consideration and the values have been chosen based on the feasibility and practicality of biomedical applications. For ultra-low-power operation, a $V_{dd}$ value of 0.5V is chosen. Using the above-mentioned constants of Table I in equation [2], the $V_{th}$ value was determined.

Table II: DESIGN GOALS

| Open-loop gain | ≥70dB |
|---|---|
| Unity Gain Bandwidth (UGB) | ≈40kHz |
| Maximum Power Dissipation | ≈100nW |
| Phase Margin | ≥70° |
| Slew rate | ≈25mV/μs |

Equation [4] was used to calculate the W/L values for each MOSFET and the drain current for each MOSFET was approximated based on the maximum tolerable power dissipation and $V_{dd}$ value. Once appropriate aspect ratios were determined for all the MOSFETs, the DC bias voltage $V_{REF1}$ was set based on the gain considerations after performing a DC analysis on the first stage of the op-amp. $V_{REF2}$ was subsequently set to meet the desired gain of the overall amplifier.

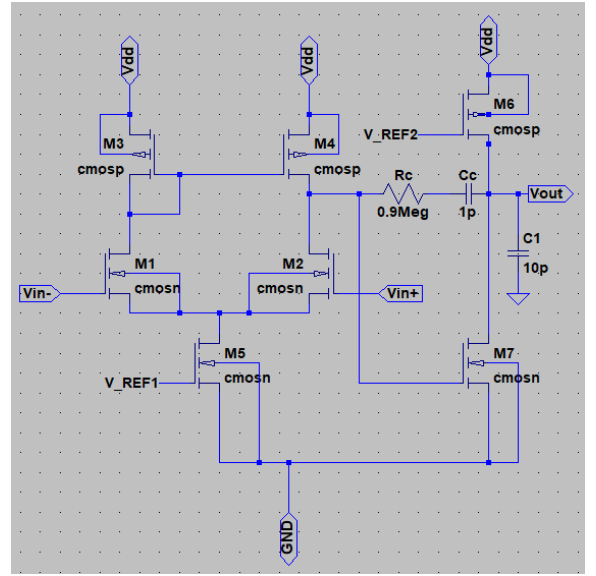

Fig. 2. Schematic of the 180nm two-stage ultra-low-power and high gain OpAmp with subthreshold biasing and Miller compensation

### C. Performance improvement and technology scaling

Once the design topology, biasing and aspect ratios were fixed, practical approaches to increase the OpAmp's performance were considered. First, the phase compensation capacitance was found using the following well-known equation to achieve the desired Unity Gain Bandwidth (UGB).

$$C_C = \frac{g_{m1}}{2\pi \cdot \text{UGB}} \quad (5)$$

where $Cc$ is the compensation capacitor, $g_{m1}$ is the transconductance of $M_1$ found using Eqn. (3). Then the compensation resistance was found to stabilize the OpAmp and fix the desired open-loop phase margin while maximizing the bandwidth. From [10],

$$R_C = \frac{1}{2g_{m7}}(1 + \sqrt{1 + \frac{4g_{m7}C_L}{g_{m1}C_1 \tan(\phi_M)}}) \quad (6)$$

where $Rc$ is the compensation resistance, $g_{m7}$ is the transconductance of $M_7$, $C_L$ is load capacitance, $C_1$ is total capacitance at the output of the first stage, and $\varphi_M$ is the desired open-loop phase margin.

To scale the circuit and understand the performance at different technology nodes, two approaches of constant field and constant voltage scaling were considered. For a scaling factor of $S$, constant field scaling is more attractive in terms of performance as the parasitic capacitance and power dissipation is scaled down by a factor of $S$ and $S^2$ respectively. However, for a low voltage application (such as the proposed work), further scaling of $V_{DD}$ below 0.5V will lead to poor signal swings, making it difficult to use for a bio-amplifier application. Moreover, constant voltage scaling is usually preferred in most practical applications as peripheral and interface circuitry may require certain voltage levels which would necessitate multiple power supply and level shifters, complicating the circuit [12].

Due to this, constant voltage scaling was preferred over constant field scaling. All dimensions were reduced by the scaling factor $S$ which is 2 and 4 for 90nm and 45nm technologies respectively. Voltages were kept constant and using the methods already described in Sec. II, the circuit performance was once again optimized. Table III shows all the transistor dimensions and network elements ensuing from the design procedure, for all three technology nodes.

Table III: MOSFET DIMENSIONS AND NETWORK ELEMENTS

| Parameter | 180nm | 90nm | 45nm |
|---|---|---|---|
| $M_{1,2}$ | 155μm/2.0μm | 65μm/1μm | 40μm/0.5μm |
| $M_{3,4}$ | 105μm/0.4μm | 1μm/0.2μm | 25μm/0.1μm |
| $M_5$ | 60μm/1μm | 30μm/0.5μm | 12.5μm/.25μm |
| $M_6$ | 95μm/10.3μm | 30μm/5.15μm | 25μm/2.575μm |
| $M_7$ | 5μm/1μm | 0.5μm/0.5μm | 1.4μm/.25μm |
| $R_C$ | 0.9MΩ | 0.15MΩ | 0.101MΩ |
| $C_C$ | 1pF | 1.45pF | 1.81pF |
| $C_L$ | 10pF | 10pF | 10pF |
| $I_{D1,2}$ | 7.423nA | 11.512nA | 14.228nA |
| $I_{D6,7}$ | 200.9nA | 240.87nA | 252.29nA |

## III. SIMULATION RESULTS AND ANALYSIS

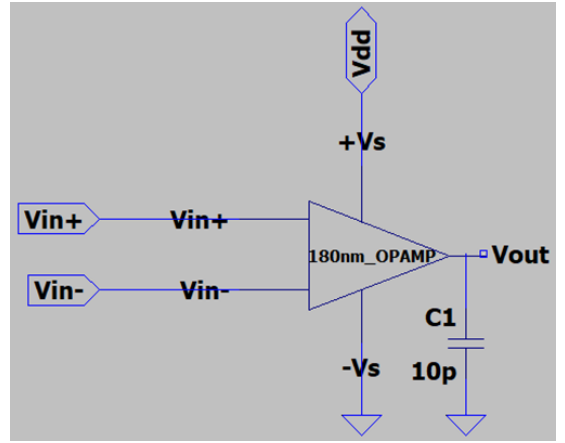

Fig. 3. Schematic of the test bench circuit for 180nm OpAmp

Figure 3 shows the schematic of the test bench circuit for the OpAmp. It consists of the OpAmp symbol with capacitive load and differential input of 4mV peak to peak and frequency 10Hz, to imitate typical biomedical signal inputs.

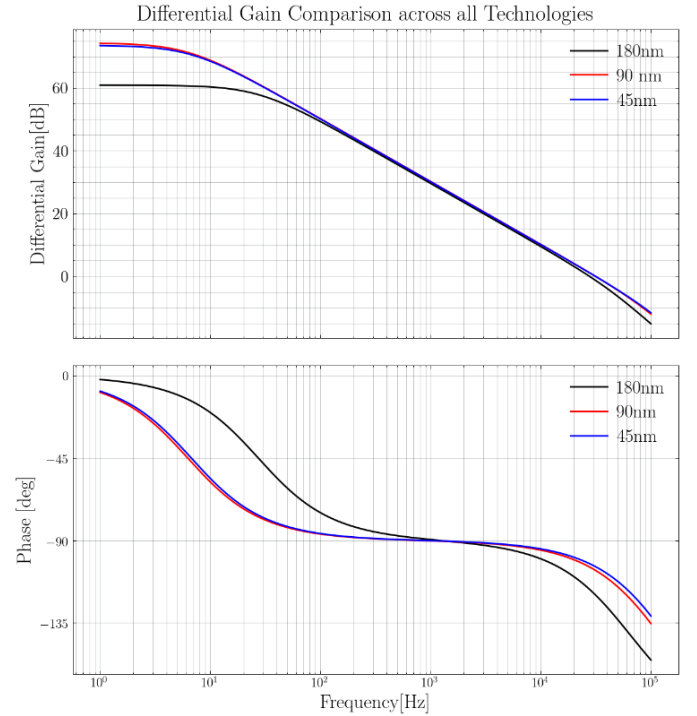

Fig. 4. AC analysis of the proposed two-stage OpAmp for 180nm, 90 nm, and 45 nm technology nodes

Figure 4 shows the open-loop amplifier frequency response for all three technology nodes. The dc gain and phase margin for the 180 nm technology node were found to be 60.86dB and 63.5° respectively as shown in Figure 4 (black curve). By scaling down the node technology to 90 nm and 45nm, the values of dc gain and the phase margin were found to be increasing, with 45nm technology showing the best performance, as shown in Figure 4. The maximum phase margin of the system was found to be 76° which ensures stable operation.

Figure 5 shows the simulated input-referred noise voltages for all three technologies. It can be observed that the flicker noise contribution becomes insignificant above roughly 1kHz. This is a relatively low corner frequency compared to 1MHz in the strong inversion design of [13] which is another advantage of subthreshold operation due to $I_D^2$ dependence of 1/f noise power [6]. The simulated values of input-referred noise at the operating frequency (10Hz) are found to be (in μV/√Hz) 0.507, 0.916, 0.492 for 180nm, 90nm, and 45nm nodes respectively.

Table IV: SIMULATION RESULTS

| Parameter | 180nm | 90nm | 45nm |
|---|---|---|---|
| DC Gain | 60.86 dB | 74.21 dB | 73.48dB |
| Phase Margin | 63.5° | 73.58° | 76° |
| $V_{dd}$ | 0.5V | 0.5V | 0.5V |
| Power Cons. | 107.88nW | 131.83nW | 140.3nW |
| Slew Rate | 27.71mV/μs | 22.28mV/μs | 33.17mV/μs |
| Unity GBW | 27.54kHz | 31.62kHz | 31.34kHz |
| Input referred noise (μV/√Hz) | 0.510 | 0.916 | 0.492 |
| $C_L$ | 10pF | 10pF | 10pF |

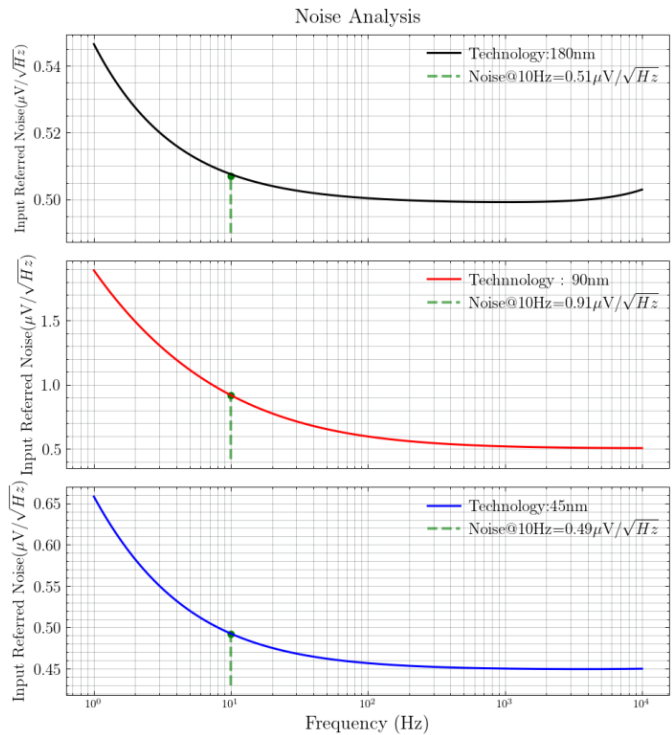

Fig. 5. Simulated input-referred noise voltages of designed OpAmp across three technologies

The measured simulation results for each technology are shown in Table IV. The value of power consumption was found to be increasing with technology scaling as expected for constant voltage scaling. It can be seen that most of the parameters meet the design goals of Table II, with a slight deviation due to unavoidable uncertainty and approximations in the hand calculations. All three technologies give high gain and ultra-low power consumption, making it ideal for a low slew rate, low-frequency applications in biomedical frontends [14].

Table V: COMPARISON OF THE PROPOSED OPAMP WITH OTHER EXISTING OPAMPS

| Parameter | Ref. [6] | Ref. [7] | Ref. [10] | Ref. [14] | This Work (90nm) | This Work (45nm) |
|---|---|---|---|---|---|---|
| Gain(dB) | 70 | 45 | 40 | 69.4 | **74.21** | 73.48 |
| Phase Margin(°) | 55 | - | 72 | 65.1 | **73.58** | 76 |
| Power Cons.(μW) | 0.075 | 6.25 | 0.112 | 0.548 | **0.131** | 0.140 |
| Input referred noise (μV/√Hz) | 0.310 | 1.8 | - | 0.29 | **0.916** | 0.492 |
| Slew rate(mV/μs) | 3 | - | 5 | 14.6 | **22.28** | 33.17 |
| $V_{dd}$(V) | 0.5 | 1.8 | 0.8 | 0.6 | **0.5** | 0.5 |
| UGB(kHz) | 18 | 2.9 | 114 | 11.35 | **31.62** | 31.34 |
| Technology (nm) | 180 | 180 | 180 | 350 | **90** | 45 |

Table V shows the comparison of the proposed OpAmp results with the other existing amplifiers reported in the literature. It is noticed that in the work reported in Ref. 10, the OpAmp achieves the UGB of 114 kHz but with a compromise on the value of gain which is 40dB. The proposed OpAmp offers better results for most of the specifications in comparison with previous work.

IV. CONCLUSIONS

In this work, a design methodology of ultra-low-power and high gain subthreshold op-amp was presented with a supply voltage of 0.5V. A comparative analysis of the proposed OpAmp was shown for the 180nm, 90nm, and 45 nm technology to analyze the performance with scaling and enable smaller yet reliable circuits. The theoretical trends have been followed, with increasing gain, phase margin, and power as the technology was scaled. Based on the analysis of these three technology nodes, it was found that the 90nm technology provides the best gain (74.21dB) and UGB (31.62 kHz) whereas the 45nm technology provides the best phase margin (76°). The proposed design showed an improvement in slew rate and UGB as compared to the previously reported results while maintaining moderately high gain (around 70 dB) and ultra-low-power operation (around 100nW) across all the technologies. This makes it particularly suitable for biomedical applications such as implantable neuroelectronic interfaces for cortical sensing.